\documentclass[amsmath,amssymb,twocolumn,amsfonts,twocolumn]{revtex4}
\usepackage{graphicx}

\def \beq {\begin{equation}}
\def \eeq {\end{equation}}
\def \tr {\rm Tr}

\begin{document}
\title{Photon statistics as an experimental test discriminating between theories of spin-selective radical-ion-pair reactions}
\author{A. T. Dellis and I. K. Kominis}

\affiliation{Department of Physics, University of Crete, Heraklion
71103, Greece}

\begin{abstract}
Radical-ion-pair reactions were recently shown to represent a rich biophysical laboratory for the application of quantum measurement theory methods and concepts. We here propose a concrete experimental test that can clearly discriminate among the fundamental master equations currently attempting to describe the quantum dynamics of these reactions. The proposed measurement based on photon statistics of fluorescing radical pairs is shown to be model-independent and capable of elucidating the singlet-triplet decoherence inherent in the radical-ion-pair recombination process.
\end{abstract}

\maketitle
\section{Introduction}
Spin-selective radical-ion-pair reactions represent a rich biophysical/biochemical system in which spin degrees of freedom can dramatically influence the fate of biologically significant chemical reactions. The study of radical-ion-pair reactions is at the core of spin chemistry \cite{steiner}, by now a mature research field directly related to photochemistry \cite{turro} and photosynthesis \cite{blankenship}. Radical-ion-pair reactions determine the late-stage dynamics in photosynthetic reaction centers \cite{boxer1,boxer2}, and furthermore there is increasing evidence that radical-ion-pair reactions underlie the avian compass mechanism, i.e. the biochemical compass used by migratory birds to navigate through the geomagnetic field \cite{schulten1,ww1,ritz,ww2,schulten2,maeda,rodgers}. Hence the fundamental understanding of these reactions if of high scientific interest.

Radical-ion pairs are molecular ions created by a charge transfer from a
photo-excited D$^*$A donor-acceptor molecular dyad DA, schematically described by the reaction ${\rm DA}\rightarrow {\rm D^{*}A}\rightarrow {\rm D}^{\bullet +}{\rm A}^{\bullet -}$, where the two dots represent the two unpaired electrons. The magnetic nuclei of the donor and acceptor molecules couple to the two electrons via the hyperfine interaction, leading to singlet-triplet mixing, i.e. a coherent oscillation of the spin state of the electrons. The reaction is terminated by the reverse charge transfer, resulting to the charge recombination of the radical-ion-pair and the formation of the neutral reaction products. It is angular momentum conservation at this step that empowers the molecule's spin degrees of freedom to determine the reaction's fate: only singlet state radical-ion pairs can recombine to reform the neutral DA molecules, whereas triplet radical-ion pairs recombine to a different metastable triplet neutral product.

The fundamental quantum dynamics of radical-ion-pair (RP) reactions rests on a master equation satisfied by $\rho$, the density matrix describing the spin state of the molecule's two electrons and magnetic nuclei. This master equation has to describe (i) the unitary evolution of $\rho$ due to the magnetic interactions within the radical-ion pairs, which is straightforward,  (ii) the loss of radical-ion pairs due to the recombination reaction leading to the creation of neutral products and (iii) the state change of unrecombined radical-ion pairs. The perplexity of the combined presence of all those phenomena is partly the reason behind the ongoing debate on the particular form of this master equation. The current standing of this debate is the following. Kominis derived \cite{komPRE} a master equation for the term (iii) and put forward a master equation \cite{komDS} taking into account the reaction term (ii). Another master equation was put forward by Jones and Hore \cite{JH}, while several authors \cite{shushin,purtov,ivanov} argued in favour of the traditional master equation of spin chemistry. So the same physical system is currently described by three theories. This situation is clearly unsatisfactory, and although theoretical arguments could in principle point to the fundamentally correct theory \cite{comment,reply}, the need for an experiment with discriminatory power is obvious. We will here propose exactly such an experiment. 
\section{Photon Statistics}
We will here describe the proposed measurement and explain the physics behind it in the next section. We consider radical pairs (RPs) that recombine only through one channel e.g. the singlet, and hence $k_{T}=0$. We also consider each recombination event to be 
accompanied by a photon emission through e.g. an exciplex fluorescing molecule. Let $n_t$ and $n_{t+dt}$ representing the photon counts in the time intervals $(t,t+dt)$ and $(t+dt,t+2dt)$. The stochastic variables $n_{t}$ and $n_{t+dt}$ have expectation values $N_{t}=k_{S}dt\tr\{Q_{S}\rho_{t}\}$ and $N_{t+dt}=k_{S}dt\tr\{Q_{S}\rho_{t+dt}\}$, respectively, given by the singlet recombination products during the respective time intervals. 
The probability to actually observe $n_t$ ($n_{t+dt}$) photons is given by the Poisson distribution with expectation value $N_{t}$ ($N_{t+dt}$).

We now define the stochastic variable $\delta n\equiv(n_{t+dt}-n_{t})/N_{t}$. If $n_1$ and $n_2$ are Poisson stochastic variables with mean values $N_1$ and $N_2$, respectively, then the probability that the difference $n_{1}-n_{2}=k$ follows the Skellam distribution, given by $f(k;N_{1},N_{2})=e^{-(N_{1}+N_{2})}(N_{1}/N_{2})^{k/2}I_{|k|}(2\sqrt{N_{1}N_{2}})$, where $I_{k}(x)$ is the modified Bessel function of the first kind. The mean and variance of the Skellam distribution are $\mu=N_{1}-N_{2}$ and $\sigma^{2}=N_{1}+N_{2}$. Therefore the mean and variance of $\delta n$ is $\overline{\delta n}=(N_{t+dt}-N_{t})/N_{t}$ and $\sigma^{2}_{\delta n}=(N_{t}+N_{t+dt})/N_{t}^{2}$, respectively. Since $N_{t}\approx N_{t+dt}$, the standard deviation can be simplified to $\sigma_{\delta n}=\sqrt{2/N_{t}}$. 
\begin{figure}
\includegraphics[width=8.5 cm]{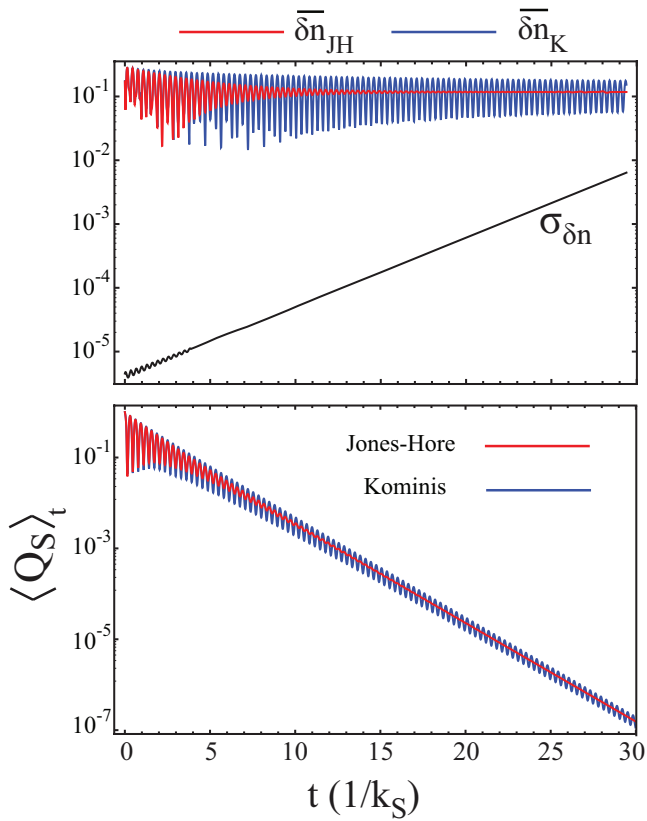}
\caption{We numerically integrate the Jones-Hore and the Kominis equation for a magnetic Hamiltonian of the form ${\cal H}=\omega_{1}s_{1z}+\omega_{2}s_{2z}$. Starting with $10^{12}$ RPs, we then calculate the expected photon counts in a time interval $\Delta t$ from $N_{t}=\int_{t}^{t+\Delta t}k_{S}\langle Q_{S}\rangle dt$, and form the ration $(N_{t}-N_{t+\Delta t})/N_{t}$, which is plotted for the two theories. We also plot the statistical error $\sigma$, almost the same for both theories.  We chose $\Delta t=1/4k_{S}$ in order to have enough statistics for $\sigma$ to be small enough to allow a statistically meaningful comparison between data and theoretical expectation.}
\label{fig1}
\end{figure}
We consider an RP with arbitrary hyperfine interactions. At high enough magnetic fields, we ignore all hyperfine interactions and keep just the $\Delta g$ contribution to S-T mixing. Thus the theoretical comparison between the two theories is completely robust and independent of the details of the molecule's hyperfine interactions. In Fig. 1a we plot the expectation value of $\overline{\delta n}$ according to the theory of Kominis ($\overline{\delta n}_{k}$) and Jones-Hore ($\overline{\delta n}_{JH}$). We also plot the standard deviation $\sigma_{\delta n}$ (roughly the same for both theories). It is clear that the oscillations in $\delta n_{K}$ decay rather slowly in time. In contrast, the oscillations in $\overline{\delta n}_{JH}$ decay fast, with a time constant on the order of the recombination rate $k_{S}$. Furthermore, the statistical error $\sigma_{\delta n}$ remains small enough for the oscillations in $\overline{\delta n}_{k}$ to be detectable for a long time well beyond $1/k_{S}$.
\section{Physical Explanation}
The expected number of photons detected in the time interval between $t$ and $t+dt$ will be $N_{t}=k_{S}\langle Q_{S}\rangle_{t}dt$, whereas those detected in the following time interval between $t+dt$ and $t+2dt$ will be $N_{t+dt}=k_{S}\langle Q_{S}\rangle_{t+dt}dt$.
Hence $\overline{\delta n}=d\langle Q_{S}\rangle_{t}/\langle Q_{S}\rangle_{t}$, where $d\langle Q_{S}\rangle_{t}=\langle Q_{S}\rangle_{t+dt}-\langle Q_{S}\rangle_{t}$. Obviously, $\overline{\delta n}$ is maximum when the slope $d\langle Q_{S}\rangle_{t}/dt$ is maximum. The maxima of the slope $d\langle Q_{S}\rangle_{t}/dt$ occur in between the peaks and troughs of $\langle Q_{S}\rangle_{t}$, which are points of minimum singlet-triplet (S-T) coherence. In other words, the maxima of  $\overline{\delta n}$ occur at instants in time when there is maximum S-T coherence. As shown in Fif.1b, where we depict the expectation value $\langle Q_{S}\rangle_{t}$ derived from the two theories, the maxima of the slope $d\langle Q_{S}\rangle_{t}/dt$ stay roughly constant for Kominis' theory, whereas they quickly decay for the Jones-Hore theory. This slope represents the "amount" of S-T coherence, and hence it is this point where the two theories fundamentally differ. We will now analyze the root of this difference. 
\subsection{Singlet-Triplet Coherence}
Multiplying $\rho$ from left and right by $1=Q_{S}+Q_{T}$, it follows that $\rho$ can be written as $\rho=\bar{\rho}+\tilde{\rho}$, where
$\bar{\rho}=Q_{S}\rho Q_{S}+Q_{T}\rho Q_{T}$ and $\tilde{\rho}=Q_{S}\rho Q_{T}+Q_{T}\rho Q_{S}$. The incoherent part of $\rho$ is represented by
$\bar{\rho}$, whereas the S-T coherent part is $\tilde{\rho}$. From the master equation for $d\rho/dt$ the rate of change $d\tilde{\rho}/dt$ can be readily calculated. 
\subsection{Prediction of the Jones-Hore Theory}
The Jones-Hore master equation is (for the considered case of having $k_{T}=0$) 
\beq
d\rho/dt=-i[{\cal H},\rho]-k_{S}(\rho-Q_{T}\rho Q_{T})\label{JHme}
\eeq
leading to
\beq
d\tilde{\rho}/dt=-k_{S}\tilde{\rho}+{\rm unitary~terms},\label{cohJH}
\eeq
where "unitary terms"=$-iQ_{S}[{\cal H},\rho]Q_{T}-iQ_{T}[{\cal H},\rho]Q_{S}$. So the coherence $\tilde{\rho}$ decays away at a rate $k_{S}$. 
\subsection{Prediction of Kominis Theory}
The master equation derived by Kominis reads for this case ($k_{T}=0$) \cite{komDS}
\begin{align}
d\rho/dt=-i[{\cal H},\rho]&-{k_{S}\over 2}\Big(Q_{S}\rho+\rho Q_{S}-2Q_{S}\rho Q_{S}\Big)\nonumber\\
&-(1-p_{\rm coh})k_{S}Q_{S}\rho Q_{S}\nonumber\\
&-p_{\rm coh}{{dn_{S}}\over {dt}}{\rho\over {\tr\{\rho\}}}\label{komME}
\end{align}
where $dn_{S}=k_{S}dt\langle Q_{S}\rangle_{t}$ and $p_{\rm coh}$ is the measure of S-T coherence introduced in \cite{komDS}.
The second, trace-preserving Lindblad term of \eqref{komME} takes into account the S-T decoherence brought about by the continuous measurement of the RP's spin state induced by the intra-molecule reservoirs \cite{komPRE,komDS}. The two last terms are the reaction terms that change $\rho$ due to the RP's that recombined into neutral products in the time interval $(t,t+dt)$. This master equation leads to 
\beq
d\tilde{\rho}/dt=-k_{S}\Big({1\over 2}+p_{\rm coh}{{\tr\{Q_{S}\rho\}}\over {\tr\{\rho\}}}\Big)\tilde{\rho}+{\rm unitary~terms}
\eeq
Neglecting for the moment the second term in the parenthesis, it is seen that the S-T coherence fades away at half the rate in this 
theory compared with the Jones-Hore theory. However, this is not the only reason for the faster decay of the amplitude of the S-T oscillations in the Jones-Hore theory evidenced in Fig.1b. 
To explain why, we rewrite the Jones-Hore master equation as
\begin{align}
d\rho/dt=-i[{\cal H},\rho]&-k_{S}\Big(Q_{S}\rho+\rho Q_{S}-2Q_{S}\rho Q_{S}\Big)\nonumber\\
&-k_{S}Q_{S}\rho Q_{S}\label{JHme2}
\end{align}
Comparing the respective terms in \eqref{JHme2} and \eqref{komME}, it is seen that (i) as noted before, the S-T decoherence takes place at double the rate in the Jones-Hore theory, and (ii) setting $p_{\rm coh}=0$ for all times in \eqref{komME} we retrieve the reaction term of the Jones-Hore theory. So the amount of spin coherence, embodied in the relative balance of $\tilde{\rho}$ versus $\bar{\rho}$ (and quantified by $p_{\rm coh}$), is not only diminished faster due to the first term in \eqref{JHme2}, but also due to the fact that the reaction term in \eqref{JHme2} distorts this balance. This point is related to the way the two theories deal with the update of the density matrix $\rho$ due to recombination, when $\rho$ is a state having large S-T coherence. This point has been analyzed in some detail in \cite{komDS}, but we will here reiterate this analysis in the present context. 

At high magnetic fields the $T_{\pm}$ states are out of resonance with the singlet, so we can assume that S-T mixing occurs just between $|S\rangle$ and $|T_{0}\rangle$. Ignoring hyperfine couplings, we effectively deal with a two-level system. Suppose that at time $t$ the spin state of the RP is the pure state $|\psi_{t}\rangle=|S\rangle$. This state has obviously zero S-T coherence. Suppose also that after the elapse of time $\tau$ the state $|S\rangle$ is transformed by the $\Delta g$ mechanism  to $|\psi_{t+\tau}\rangle=(|S\rangle|+|T_{0}\rangle)/\sqrt{2}$. The corresponding density matrices will be $\rho_{t}=|S\rangle\langle S|$ and $\rho_{t+\tau}=(|S\rangle\langle S|+|T_{0}\rangle\langle T_{0}|+|S\rangle\langle T_{0}|+|T_{0}\rangle\langle S|)/2$. Let us neglect for this discussion the change of the density matrix of the unrecombined RPs, and focus on the change of the density matrix due to the recombining RPs.
The update of $\rho_{t}$ due to RPs recombining between $t$ and $t+dt$ is the same for both theories, since this is the incoherent limit. However,
the update of $\rho_{t+\tau}$ due to RPs recombining between $t+\tau$ and $t+\tau+dt$ differs fundamentally in the two theories. In the Jones-Hore theory $\rho_{t+\tau}$ changes due to recombination by $d\rho_{r}=-k_{S}dtQ_{S}\rho_{t+\tau}Q_{S}=-k_{S}dt|S\rangle\langle S|$. In contrast, in the Kominis theory $d\rho_{r}=-dn_{S}\rho_{t+\tau}$, where $dn_{S}=k_{S}dt\langle Q_{S}\rangle_{t+\tau}$ is the number of RPs that recombined within the time interval $(t+\tau,t+\tau+dt)$. Stated qualitatively, in the theory of Kominis and in the case of maximal S-T coherence, one removes the complete density matrix $\rho_{t+\tau}$ as many times as many RPs recombined. In contrast, the Jones-Hore theory projects out of $\rho_{t+\tau}$ the singlet part $Q_{S}\rho_{t+\tau}Q_{S}$, no matter how coherent the state $\rho_{t+\tau}$ is. This has the result that the rest, surviving radical-ion pairs appear to be less coherent. In contrast, the surviving RPs in the Kominis approach have not lost any coherence at all. The artificial loss of S-T coherence due to the projective recombination term in the Jones-Hore theory is thus the main reason behind the rapid decay of the oscillations seen in Fig. 1b. 

\section{Spin Relaxation}
\begin{figure}
\includegraphics[width=8.5 cm]{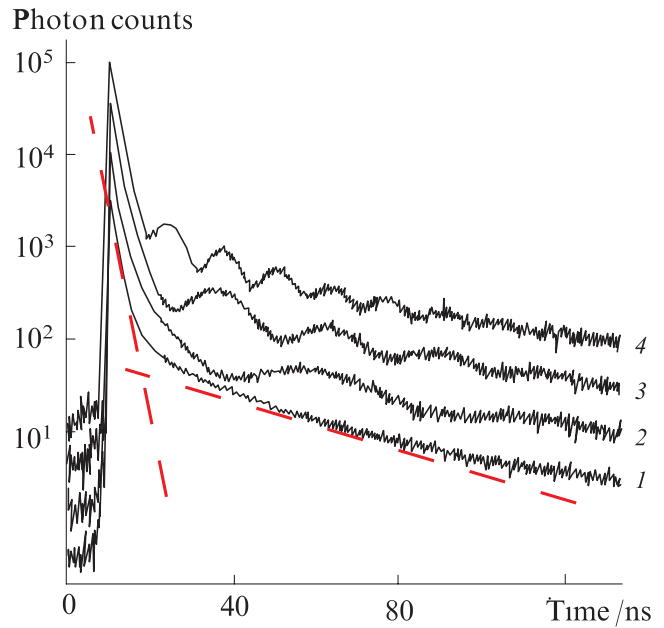}
\caption{Figure 4 of \cite{molin}.  Photon counts of singlet recombining (diphenylsulphide-d$_{10})^{+}$/(p-terphenyl-d$_{14})^{-}$ at various magnetic fields, increasing from 1 to 4. At the highest magnetic field one observes quantum beats stemming from the $\Delta g$ mixing mechanism. The "calibration" trace 1 at the lowest magnetic field is seen to be governed by two rates, a fast rate on the order of $k_{S}$ and a slow rate stemming from the strong projection produced by $k_{S}$ on the state that has oscillated to the triplet. In any case, the quantum beats decay at a rate 
much smaller than $k_{S}$, which is what the theory of Kominis predicts.}
\label{fig2}
\end{figure}
The obvious difference between the predictions of the two theories will fade away if other spin-relaxation mechanisms are dominant beyond the 
S-T decoherence process internal in the molecule. We emphasize that the latter decoherence mechanism, present in both theories, is an unavoidable one
fundamentally linked to the charge recombination process. It's physical origin has been exhaustively explained in \cite{komPRE} and \cite{komDS}. Other relaxation mechanisms, for example spin-changing collisions with other molecules, can in principle be suppressed, e.g. by performing the measurement at low enough temperatures. In any case, if the spin relaxation rate of all such other
mechanisms is denoted by $k_{sr}$, we can state that that in order for the difference between the two theories to surface, $k_{sr}$ must be significantly shorter than $k_{S}$. Otherwise, S-T coherence will be quickly suppressed, i.e. $p_{\rm coh}\rightarrow 0$, and then the two master
equations will yield qualitatively similar results, as the respective reaction terms will coincide. 
The condition $k_{sr}<k_{S}$ can be experimentally satisfied. Indeed, there already exist experimental data that provide evidence for our assertions. In \cite{molin} Molin presents photon counts of singlet recombining (diphenylsulphide-d$_{10})^{+}$/(p-terphenyl-d$_{14})^{-}$ at various magnetic fields, demonstrating the oscillations due to $\Delta g$ mixing at high fields. The data is reproduced in Fig. 2. It is evident that there exist two rates a fast rate (on the order of $k_{S}$) and a slow rate (much smaller than $k_{S}$). It is also evident that while the RP population decay is governed by the fast rate, the quantum beats decay at the slow rate. This is in contrast to the predictions of the Jones-Hore theory, according to which the amplitude of the quantum beats should decay at the rate $k_{S}$. 
\section{Conclusions}
It is important to stress that the particular measurement we propose is model-independent, in the following sense. Clearly the time evolution of 
observables like $\langle Q_{S}\rangle$ or the magnetic-field effect (MFE) are predicted to be different by the two theories. However, in practice it would be almost impossible for the measurement to discern absolute differences in $\langle Q_{S}\rangle$ or the MFE signal, since it would be possible to attribute those to an imperfect understanding of e.g. the RP's magnetic interactions. In contrast, in the measurement we propose, normalizing the photon difference by $N_{t}$ largely alleviates this problem, and the two theories predict a clearly different trend, no matter what the details of the magnetic interactions are, i.e. how many nuclear spins there are or what are the exact values of the hyperfine couplings.

\end{document}